\documentclass[reprint,superscriptaddress,aps,prd,floatfix,nofootinbib,preprintnumbers]{revtex4-2}
\usepackage{graphicx}
\usepackage[dvipsnames]{xcolor}

\usepackage{amsmath}
\usepackage{amsfonts}
\usepackage{amssymb}
\usepackage{bm}
\usepackage{slashed}
\usepackage{datetime}
\usepackage{hyperref}
\usepackage{soul} 
\usepackage{enumitem}
\usepackage{relsize}
\usepackage{blindtext}
\usepackage{tabulary}
\usepackage{natbib}
\usepackage{orcidlink}

\definecolor{bluemv}{HTML}{2a67bd}
\definecolor{darkbluemv}{HTML}{064789}
\definecolor{lightbluemv}{HTML}{80DEEA}
\definecolor{magentamv}{HTML}{EC407A}
\definecolor{orangemv}{HTML}{ff9100}
\definecolor{greenmv}{HTML}{009e60}

\hypersetup{colorlinks, citecolor=blue, linkcolor=blue, urlcolor=blue, linktocpage=true}

\makeatother   
\newcommand{\GeV}{{\rm \,GeV}}

\newcommand{\MeV}{{\rm \,MeV}}

\newcommand{\sigmav}{\langle \sigma_{ann} v \rangle}

\newcommand{\mchi}{m_\chi}

\newcommand{\Uaf}{U_{\alpha 4}}

\begin{document}

\title{Neutrino portals to MeV WIMPs with s-channel mediators}

\author{Nicole F. Bell \orcidlink{0000-0002-5805-9828}}
\email{n.bell@unimelb.edu.au}
\author{Matthew J. Dolan \orcidlink{0000-0003-3420-8718}}
\email{matthew.dolan@unimelb.edu.au}
\author{Avirup Ghosh \orcidlink{0000-0002-4781-842X} }
\email{avirup.ghosh@unimelb.edu.au}
\author{Michael Virgato \orcidlink{0000-0002-8396-0896} }
\email{mvirgato@student.unimelb.edu.au}
\affiliation{ARC Centre of Excellence for Dark Matter Particle Physics, \\
School of Physics, The University of Melbourne, Victoria 3010, Australia}

\begin{abstract}
Large-scale neutrino detectors currently under construction will have the unique ability to probe the annihilation of low-mass thermal-relic dark matter to neutrinos. This represents an essential test of the thermal freezeout paradigm. This raises the question: what viable UV-complete models are there in which dark matter annihilates dominantly to neutrinos? 
We discuss models that fulfill this criteria, and are invariant under the Standard Model gauge group, for both scalar and  fermionic dark matter. Specifically, we construct new models in which annihilation via the $s$-channel exchange of a scalar or pseudoscalar mediator achieves the correct relic density. In these models, dark matter is stabilised by an exact or a softly-broken lepton-number symmetry. The parameter space of such models will be probed, almost entirely, by the combination of JUNO, Hyper-Kamiokande and CMB-S4.
\end{abstract}

\maketitle

\section{Introduction}
\label{sec:intro}

Pinning down the identity and interactions of dark matter (DM) is the subject of many active avenues of investigation in the particle physics community. One of the most important questions we must answer is whether DM is a thermal relic, with an abundance set by DM annihilation in a thermal freezeout process.  The most direct way to probe this question in the present day universe is via indirect detection,  because observable signals are determined by the very same annihilation cross section that controls freezeout.

Indirect detection is the search for dark matter annihilation products, emanating from regions of high dark matter density, such as the Galactic Centre or dwarf galaxies. In recent years, observational searches for such DM annihilation fluxes have become increasingly sensitive. Dark Matter which annihilates to visible final states will lead to the production of photons or cosmic-rays either directly, or via the decay or hadronisation of primary annihilation products, or via energy losses during propagation.  This leads to strong limits on annihilation to almost all SM final states~\cite{Leane:2018kjk}. 

{\it Invisible} final states such as neutrinos, however, are much harder to detect, and thus the limits on the DM annihilation to neutrinos are correspondingly weaker. Yet they are of critical importance, as they represent the most robust limits on the total dark matter annihilation cross section~\cite{Beacom:2006tt,Yuksel:2007ac}. Large-scale neutrino telescopes have sensitivity to such neutrinophilic dark matter, particularly at low masses (see~\cite{Arguelles:2019ouk} for a summary). For example, data from Super-Kamiokande (SuperK) was used to set constraints in the MeV mass range via re-interpretation of searches for the diffuse flux of supernovae neutrinos and relic supernovae electron anti-neutrino flux~\cite{Arguelles:2019ouk}. Importantly, the upcoming Hyper-Kamiokande (HyperK) experiment~\cite{Hyper-Kamiokande:2018ofw} will be able to probe DM annihilation to neutrinos, with a thermal relic cross section, over an extended MeV-scale mass range~\cite{Bell:2020rkw}.

With observational probes of sub-GeV DM annihilation to neutrinos finally at our disposal, it is important to thoroughly explore the model space which can yield such a signal. Specifically: what viable models are there in which DM annihilates dominantly to neutrinos, with  zero (or suppressed) branching ratios to charged leptons or other more detectable SM final states? In order to yield a detectable indirect detection signal, we also require the annihilation to be $s$-wave dominated. We shall construct new models that achieve this goal, to add to those models which already exist~\cite{Cherry:2014xra, Blennow:2019fhy, Coito:2022kif, Bertoni:2014mva, Batell:2017cmf, Okawa:2020jea, Iguro:2022tmr}. These models all use a neutrino portal structure, in which the DM couples to sterile neutrinos that mix with the active neutrinos.

Our models involve scalar and fermionic SM-singlet sub-GeV thermal dark matter. In both cases the DM is charged under a global $U(1)_L$ symmetry, which ensures the DM is stable on cosmological time-scales. The presence of this global $U(1)_L$ allows the 
DM to couple to the extended neutrino sector of the SM via a scalar charged under $U(1)_L$. This allows the DM to annihilate into SM neutrinos via $s$-channel scalar-mediated processes. We show that these models are excellent targets for future neutrino detectors, and that much of the parameter space can be probed either by HyperK searches for the annihilation of Galactic DM or CMB-S4 constraints on the effective number of neutrino degrees of freedom in the early universe.

The paper is organized as follows: In Section~\ref{sec:model_build} we discuss the subtleties involved in constructing models with significant active-sterile mixing. In Section~\ref{sec:models}, we discuss two simple neutrino-portal models for sub-GeV scalar and fermionic DM, and analyse present experimental constraints and sensitivities of several future observations. Our conclusions can be found in Section~\ref{sec:conclusion}, and further details are provided in Appendices~\ref{app:Neutrinos} and \ref{app:Annxsec_Thermav}.

\section{Active-sterile neutrino mixing}
\label{sec:model_build}

For dark matter to annihilate into neutrinos, there are two possibilities~\footnote{A 
collection of interaction structures that can 
achieve this along with neutrino mass generation are catalogued in~\cite{Lindner:2010rr}. However, not all the interactions they considered can be embedded in a UV-complete scenario.}. A dark-sector field must couple either directly to the SM lepton doublet, and hence to active neutrinos or, alternatively, to SM singlet right-handed (sterile) neutrinos that mix with the active neutrinos. The latter are known as \textit{neutrino portal} models.

A direct coupling to the SM lepton doublet requires new fields in non-trivial representations of the SM gauge group, which can be either the DM itself or a dark-sector mediator. Either way, there exists strong constraints from collider searches on these scenarios, although some parameter space for light DM is still viable~\cite{Okawa:2020jea, Iguro:2022tmr}.

Consequently, we opt to focus on SM singlet dark matter which annihilates via the neutrino portal. In these models, a SM singlet right-handed (RH) sterile neutrino $N_R$ is added which couples to the left-handed (LH) SM lepton doublet $l_L$ via the portal operator
\begin{equation}
\label{eq:nu_portal}
y_\nu \overline{l_L} \tilde{H} N_R,
\end{equation}
where $y_\nu$ is the neutrino Yukawa coupling and $H\;( \tilde{H} = i \sigma_2 H^*)$ is the SM Higgs doublet.
In the presence of this operator alone, the neutrinos obtain Dirac masses, and no active-sterile mixing exists. We could achieve active-sterile mixing via the introduction of a Majorana mass term for $N_R$ to permit the usual type-I seesaw; however, this would result in very tiny active-sterile mixing angles, too small to be of phenomenological interest, as per the conclusions of ref.~\cite{Lindner:2010rr}.

Instead of the usual type-I/II/III see-saw models, we shall instead consider neutrino sectors similar to the inverse~\cite{Mohapatra:1986aw, Nandi:1985uh, Mohapatra:1986bd} and linear~\cite{Wyler:140959, Akhmedov:1995ip, Akhmedov:1995vm, Malinsky:2005bi} seesaw mechanisms, as considered in refs.~\cite{Bertoni:2014mva,Batell:2017cmf, Blennow:2019fhy}.  
The advantage of these alternative seesaw structures is that they permit sizeable active-sterile mixing angles, even with electroweak scale Majorana masses. In these models, a (pseudo) Dirac neutrino, $n_D$, of mass $m_N$, is formed from the RH field $N_R$ and a LH field $S^c_R$. The $N_R$ field couples to the SM lepton doublet as per the neutrino portal operator of eq.~\ref{eq:nu_portal}, while $S_R$ will allow our DM candidate to annihilate dominantly into the SM neutrinos. The neutrino sector Lagrangian can thus can be expressed as: 
\begin{equation}
    -\mathcal{L}_\nu = m_N \overline{S^c_R}N_R + y_\nu \overline{l_L}\tilde{H}N_R + h.c.
\label{eq:lagnu}
\end{equation}
where the Weyl fields $N_R$ and $S_R$ carry conserved lepton numbers (i.e., $U(1)_L$ charges) $+1$ and $-1$, respectively. 

After Electroweak Symmetry Breaking (EWSB), eq.~\ref{eq:lagnu} gives rise to a $5\times 5$ neutrino mass matrix that, upon diagonalization, results in three massless Majorana states $n_{L\,i}$ (with $i=1,2,3$) and one massive Dirac state $n_D$ (see Appendix~\ref{app:Neutrinos} for details). The generation of non-zero light neutrino masses can be achieved by introducing a small lepton number violating Majorana mass term for $S_R$ (inverse see-saw) or a mass term coupling involving $S_R$ and $l_L$ (linear see-saw). The inclusion of these $\Delta L = 2$ mass terms is not expected to significantly modify the dark matter phenomenology we study here so, for simplicity, we will not introduce them. In this scenario, the mixing angles obey the same expressions as for a canonical seesaw model, but $m_N$ and $y_\nu$ are now essentially free parameters, allowing for significantly larger mixing angles.

Although these mass/mixing frameworks permit large active-sterile mixing, there exist several important experimental constraints.
For sterile neutrinos of mass
$\mathcal{O}({\rm MeV}) - \mathcal{O}(100\,{\rm GeV})$, collider searches, cosmological and astrophysical observations have set stringent upper bounds on the active-sterile mixing angles, in 
the range $|U_{\alpha\,4}|^2 \lesssim  10^{-12} - 10^{-5}$~\cite{Boyarsky:2009ix, Ruchayskiy:2012si,Vincent:2014rja,Bolton:2022pyf}.
For sterile neutrinos of mass $\gtrsim 108\,{\rm GeV}$, however, the only relevant constraints arise from the non-unitarity of the light neutrino mixing matrix, which alters the Electroweak Precision Observables (EWPO)~\cite{Bolton:2022pyf}. We thus take the sterile neutrino mass to be $\gtrsim 108\GeV$, so that, only the EWPO constraint imposes an upper limit of $\sim 10^{-2}$ on the active-sterile mixing $|U_{\alpha\,4}|$. This, in turn, imposes an upper-limit on the DM-neutrino interaction strength for any given DM mass, as we discuss in the following section.

\section{Neutrino Portal DM Models}
\label{sec:models}

Having identified a framework that permits reasonably large active-sterile mixing, we now introduce a mediator which couples the DM to the sterile neutrinos. In Table~\ref{tab:DM_models} we provide a classification of minimal neutrino-portal models for various mediator choices. We have included only those models that permit $s$-wave annihilation of DM to SM neutrinos. This explains the absence of some particular DM-mediator combinations (such as scalar DM with a vector mediator), for which the annihilation cross-section has no $s$-wave piece. The table contains existing models that ultilise a low-scale seesaw mechanism, together with the new models that will be presented here~\footnote{Note that one can construct more involved models which meet our criteria, but do not fit into this simplistic classification scheme, e.g., the semi-annihilating scalar DM model of ref.~\cite{Miyagi:2022gvy}.}. These are $s$-channel (pseudo)-scalar mediated neutrino portal models, for sub-GeV scalar or fermionic DM. 

\begin{table}[t]
    \centering
    \begin{tabular}{c c c c}
        \textbf{Dark Matter} & \textbf{Channel} & \textbf{Mediator} & \textbf{Refs.}\\\hline\hline
         Fermion & $s$ & Pseudoscalar & This work\\
         Fermion & $s$ & Vector & \cite{Cherry:2014xra, Blennow:2019fhy, Coito:2022kif,Foldenauer:2018zrz,Asai:2020qlp}\\
         Fermion & $t$ & Scalar & \cite{Bertoni:2014mva, Batell:2017cmf, Blennow:2019fhy, Okawa:2020jea, Coito:2022kif,  Iguro:2022tmr} \\
         \hline
         Scalar  & $s$ & Scalar & This work\\
         Scalar  & $t$ & Fermion & \cite{Bertoni:2014mva, Batell:2017cmf, Blennow:2019fhy, Okawa:2020jea, Iguro:2022tmr} \\\hline
    \end{tabular}
\caption{Classification of the minimal models which 
permit a sub-GeV DM candidate annihilating into active 
neutrinos via $s$-wave processes. Whenever available, 
the existing models in the literature have been 
pointed out.}
\label{tab:DM_models}
\end{table}

The $s$-channel mediator, $\Phi$, in our models, will be an SM singlet. If $\Phi$ is also a $U(1)_L$-singlet, the interaction 
$\left(\Phi \overline{S_R}\,N^c_R + h.c. \right)$ is allowed. This would lead to the pseudo-Dirac state $n_D$ being produced in DM annihilations as  $N_R$ does not couple to light neutrinos (see Appendix~\ref{app:Neutrinos}). Since we focus on MeV-scale dark matter and restrict the $n_D$ mass to be greater than 108~GeV, this case is not relevant for our work.

Therefore, we allow the mediator $\Phi$  to couple independently to $N_R$ and $S_R$. This implies the DM can annihilate into light neutrinos via the term $\Phi\,\overline{S^c_R}\,S_R$, without requiring the sterile neutrinos to be light. Such an interaction term requires $\Phi$ to carry a lepton number $+2$. That, in turn, implies that the DM will have a non-zero lepton number (unless lepton number is explicitly violated by two units, which we do not consider).

\subsection{Scalar Dark Matter}
\label{ssec:scalar_DM}

We first discuss a model where the DM candidate is a complex scalar, $\rho$, that carries a $U(1)_L$ charge of +1. The mediator $\Phi$ is a SM-singlet complex scalar field with $U(1)_L$ charge of +2, which couples to both the DM $\rho$ and the sterile state $S_R$. The $U(1)_L$ charge assignments of the SM singlet fields are summarized in Table~\ref{tab:scalar_qnums}.

\begin{table}[htb!] 
    \centering
    \begin{tabular}{|c|c|c|c|c|}
    \hline
                 & $\rho$ & $\Phi$ & $N_R$  & $S_R$\\\hline
         $U(1)_L$        & $+1$   & $+2$   & $+1$ & $-1$ \\
         \hline
    \end{tabular}
    \caption{$U(1)_L$ charges of the SM singlet scalars and fermionic fields in the scalar DM model.}
    \label{tab:scalar_qnums}
\end{table}

\subsubsection{Overview of the model}

Given the lepton number assignments of the SM singlet scalars and the fermionic fields, the $U(1)_L$ invariant Lagrangian is given by:
\begin{align}
    -\mathcal{L} & = - \mathcal{L}_\nu + \left(\frac{\lambda_{S\phi}}{2}\overline{S^c_R}S_R\,\Phi + \frac{\lambda_{N\phi}}{2}\overline{N_R}N^c_R\Phi + h.c.\right)\nonumber\\ 
    & + V(H,\Phi,\rho),
    \label{eq:lag_scalar}
\end{align}
where the scalar potential is,
\begin{align}
V(H,\Phi,\rho) &= \mu_\rho^2|\rho|^2 + \mu_\phi^2|\Phi|^2  - \mu_H^2|H|^2\nonumber\\ 
    & + g_\rho |\rho|^4 + g_\phi |\Phi|^4 + g_H |H|^4\nonumber\\
    & + g_{\phi H} |\Phi|^2 |H|^2 + g_{\rho H}  |\rho|^2 |H|^2\nonumber\\
    & + g_{\rho \phi} |\rho|^2 |\Phi|^2
    + \frac{\mu_{\rho\phi}}{2}\left(\rho^2 \Phi^* + h.c.\right),
    \label{eq:pot_scalar}
\end{align}
and $\mathcal{L}_\nu$ is same as in eq.~\ref{eq:lagnu}. We assume $\mu^2_\rho,\mu^2_\Phi,\mu^2_H > 0$, and hence the global $U(1)_L$ symmetry remains unbroken, which explains the stability of the DM $\rho$~\footnote{We shall not consider the possibility of gravity-mediated global symmetry breaking~\cite{Mambrini:2015sia}, which is usually Planck-scale suppressed.}. 

Once the electroweak symmetry is broken, both the DM $\rho$ and the mediator $\Phi$ couple to the SM Higgs field $H$ via the quartic couplings $g_{\rho H}$ and $g_{\phi H}$. This has two main consequences: (i) the DM and mediator masses receive tree-level contributions from the Higgs vev $v_H$; and (ii) we open DM annihilation channels to light charged leptons ($e$ and $\mu$) via $s$-channel Higgs exchange. 

The DM and mediator masses satisfy
\begin{align}
    m_\rho^2 &= \mu_\rho^2 + \frac{1}{2} g_{\rho H} v_H^2, \\
    m_\Phi^2 &= \mu_\phi^2 + \frac{1}{2} g_{\phi H} v_H^2,
\end{align}
respectively. Since we want both of these masses to be $\mathcal{O}({\rm MeV})$, both $g_{\rho H}$ and $g_{\phi H}$ must be small. Requiring the higgs contributions to be less than $10\%$ of the actual masses of the DM and the mediator, we obtain,
\begin{eqnarray}
\label{eq:grH_mass_constr}
g_{\rho H} &\lesssim & 3.3\times 10^{-11} \left(\frac{m_\rho}{10 \MeV}\right)^2,\\ 
g_{\phi H} &\lesssim & 3.0\times 10^{-10} \left(\frac{m_\Phi}{30 \MeV}\right)^2. 
\label{eq:gpH_mass_constr}
\end{eqnarray}
Although there exists no obvious naturalness argument for the smallness of $g_{\rho H}$ and $g_{\phi H}$, we note that, setting these parameters small at tree-level (along with $g_{\rho \phi}$) will ensure their smallness at all perturbative orders. Moreover, such tiny scalar quartic couplings are often encountered in the models of scalar singlet freeze-in dark matter (e.g., see~\cite{McDonald:2001vt,Yaguna:2011qn}).

Invisible Higgs decays can also provide constraints on $g_{\rho H}$ and $g_{\phi H}$. This decay rate is given by
\begin{align}
    \Gamma(H\rightarrow \rm{invis.}) & = \frac{v_H^2}{64 \pi m_H} \sum_{i = \rho, \Phi} g_{i H}^2 \sqrt{1 - \frac{4m_i^2}{m_H^2}} \\
    & \simeq  \frac{v_H^2}{64 \pi m_H} \left( g_{\rho H}^2 + g_{\phi H}^2\right)\,{\rm for}\,\,m_{\rho,\Phi} \ll m_H.
\end{align}
The invisible decay branching fraction is constrained to 
be $<19\%$ of the total decay width of $H$~\cite{CMS:2018yfx}, 
which leads to the following constraint
\begin{equation}
g_{\rho H}, g_{\phi H} < 1.3 \times 10^{-2}.
\end{equation}
For the parameter space we are interested in, the naturalness constraints of eqs.~\ref{eq:grH_mass_constr} and \ref{eq:gpH_mass_constr} are always the more stringent ones.

\subsubsection{DM annihilation cross-sections and relic density}
\label{sssec:scalar_relic_abundance}

\begin{figure*}[htb!]
    \centering
    \includegraphics[width=8.9cm,height=6.0cm]{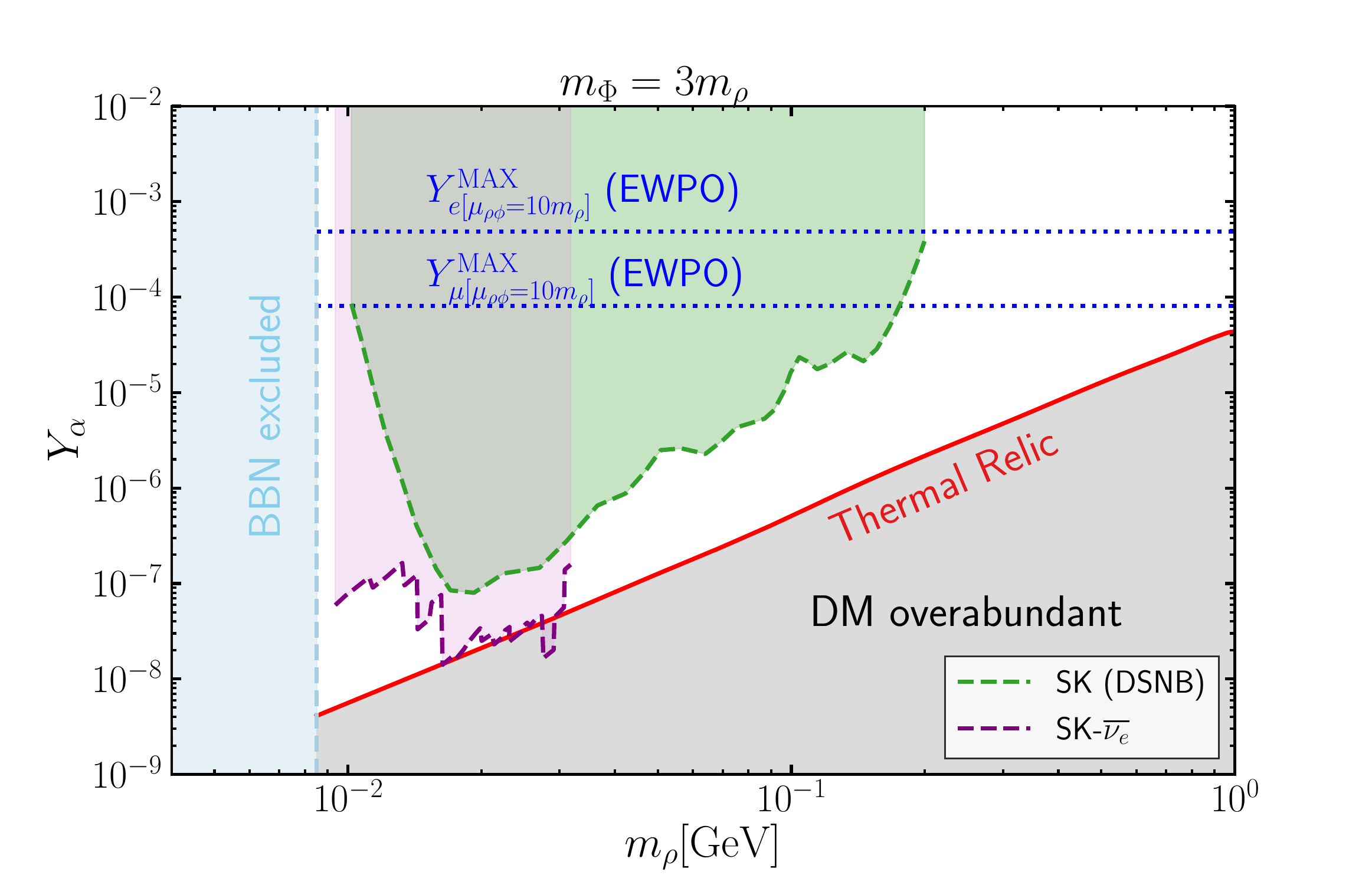}
    \includegraphics[width=8.9cm,height=6.0cm]{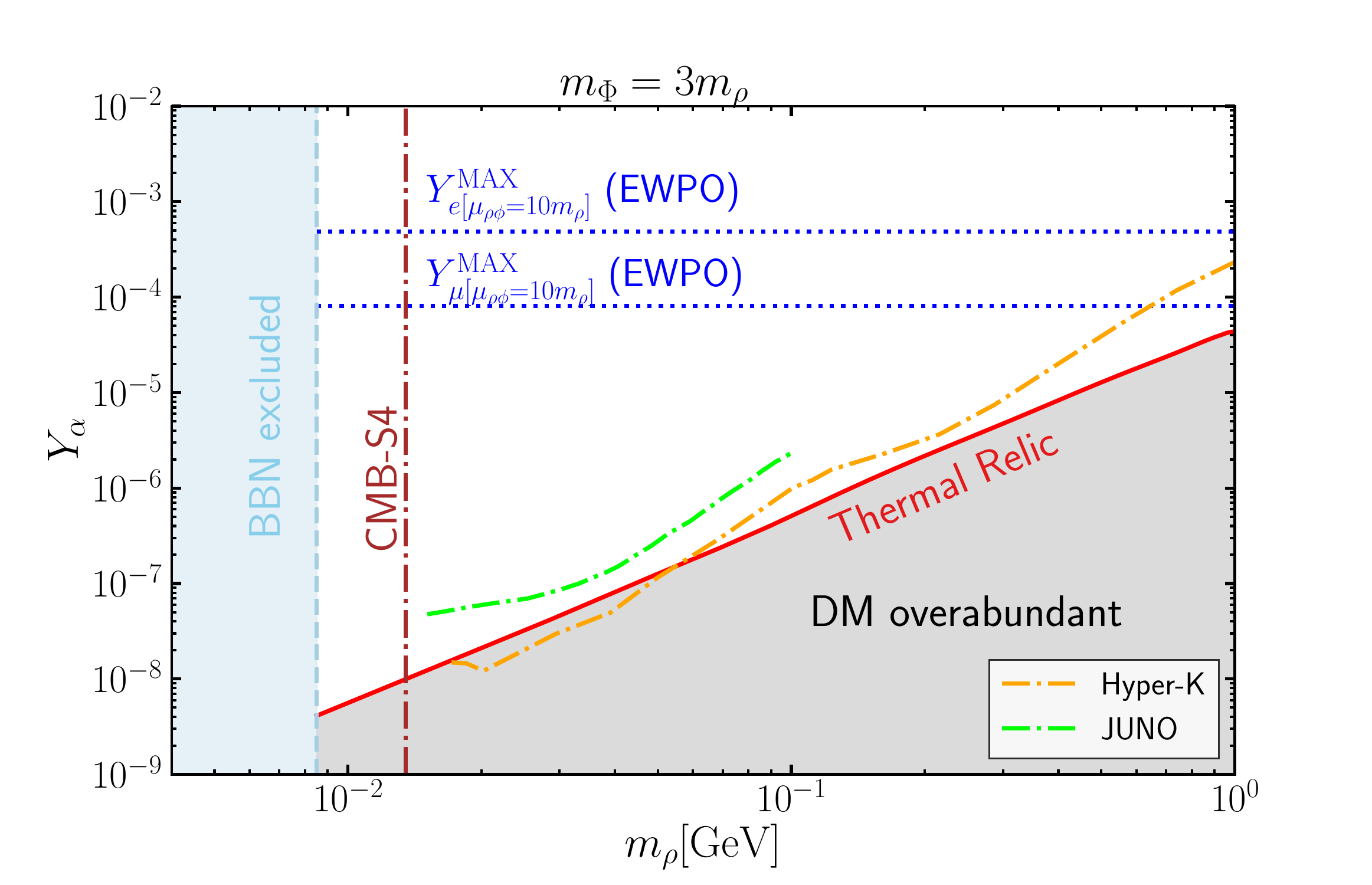}
    \caption{Parameter space for the scalar DM model with $m_\Phi = 3 m_\rho$. The red lines show values of $Y_\alpha$ which lead to the correct DM relic abundance. In the gray-shaded regions below these lines, the DM is overproduced. The existing constraints from the Super-K (green and purple dashed lines) are shown in the left panels, while the light blue shaded regions indicate BBN constraints. In the right panels, the projected sensitivities of the Hyper-K (orange) and JUNO (light green) are shown along with the CMB-S4 projected sensitivity (brown dot-dashed line). The horizontal blue dotted lines in each panel represent the maximum allowed value of $Y_\alpha (\alpha = e,\mu)$ from the non-unitarity of the light neutrino mixing matrix, for the particular parameter choices indicated. See the text for details.} 
    \label{fig:CS_s_chann_Ya}
\end{figure*}

In order to calculate the DM relic density we need the 
DM annihilation cross-section to active neutrinos. In 
this model, DM-mediator vertex is determined by the term 
$\frac{\mu_{\rho\phi}}{2}\left(\rho^2 \Phi^* + h.c. \right)$, 
while the neutrino-mediator vertex originates from 
$\frac{\lambda_{S\phi}}{2}\overline{S^c_R}S_R\,\Phi$ 
(see appendix~\ref{app:Neutrinos}, for details). 
These give the following annihilation cross-section 
for the process $\rho \rho \rightarrow \nu\nu$ 
(see appendix~\ref{app:Annxsec_Thermav}): 
\begin{align}
     \sigma_{ann}(\rho \rho \rightarrow \nu\nu) = 
     & \frac{(\mu_{\rho \phi } \lambda_{S\phi})^2 }{128\pi}  \left(\sum_\alpha |U_{\alpha 4}|^2\right)^2 \nonumber\\ 
     & \times \frac{1}{(s - m_\Phi^2)^2}\sqrt{\frac{s}{s - 4m_\rho^2}},
\end{align}
which has the following $s$-wave piece,
\begin{align}
     \sigmav(\rho \rho \rightarrow \nu\nu) = \frac{(\mu_{\rho \phi} \lambda_{S\phi})^2}{64\pi\left(m^2_\Phi - 4m^2_\rho\right)^2}
     \left(\sum_\alpha |U_{\alpha 4}|^2\right)^2.
\label{eq:scalarDM_swaveA}
\end{align}

The presence of the $g_{\rho H}$ coupling induces DM annihilation to light charged leptons (e.g., $e^\pm$, $\mu^\pm$) via $s$-channel Higgs exchange. The cross-section is
\begin{align}
\sigma_{ann}(\rho \rho^* \rightarrow l^+ l^-) &= \frac{g_{\rho H}^2 m_l^2}{128\pi s}\sqrt{\frac{s - 4 m_l^2}{s - 4 m_\rho^2}}\frac{s - 4 m_l^2}{(s - m_H^2)^2},
\end{align}
with the corresponding $s$-wave component being, 
\begin{align}
    \sigmav(\rho \rho^* \rightarrow l^+ l^-) & = \frac{g_{\rho H}^2 m_l^2}{64\pi m_\rho^3 }\frac{(m_\rho^2 - m_l^2)^{3/2}}{(m_H^2 - 4 m_\rho^2)^2}\nonumber\\ 
    & \approx \frac{g_{\rho H}^2 m_l^2}{64\pi m_H^4 }\left( 1 - \frac{ m_l^2}{m_\rho^2}\right)^{3/2},
\label{eq:svannee_swave}    
\end{align}
where $l=e,\mu$.
In the last line of eq.~\ref{eq:svannee_swave}, we have used $m_H\gg m_\rho$. Due to the smallness of $g_{\rho H}$, DM annihilation to neutrinos always dominates over annihilation to charged leptons, and hence dictates the relic abundance.

To simplify our calculations, we shall consider the case where the sterile neutrinos mix with only one flavour (i.e., $\alpha = e$ or $\mu$) at a time, and define the following combination of free parameters 
\begin{equation}
Y_{\alpha} = \left(\frac{\mu_{\rho \phi}}{m_\rho} \lambda_{S\phi} |U_{\alpha 4}|^2\right)^2.    
\end{equation} 
The $s$-wave component for DM annihilation to neutrinos can be expressed in terms of $Y_\alpha$ as 
\begin{eqnarray}
\sigmav(\rho \rho \rightarrow \nu_\alpha \nu_\alpha) = \frac{Y_\alpha}{64\pi}\frac{m^2_\rho}{\left(m^2_\Phi - 4\,m^2_\rho\right)^2}.
\label{eq:scalarDM_swave}
\end{eqnarray}
Since the DM $\rho$ is a complex scalar, the DM $\rho$ and the anti-DM $\rho^*$ will contribute equally to the DM relic density $\Omega_{\rm DM}$ (given the absence of any CP-violating couplings). Therefore, the annihilation cross-section in eq.~\ref{eq:scalarDM_swave} must give rise to a DM density $\Omega_{\rm DM}/2$. 

We have implemented this model in 
\textsc{FeynRules}$\_$v2~\cite{Christensen:2008py, Alloul:2013bka}, and interfaced the output with \textsc{micrOMEGAs}$\_$v6~\cite{Belanger:2001fz,Alguero:2023zol} 
to calculate the relic abundances. Fig.~\ref{fig:CS_s_chann_Ya} shows the values of $Y_\alpha$ that give rise to the correct relic density (solid red lines) as functions of the DM mass $m_\rho$. 
We have assumed $m_\Phi/m_\rho = 3$, so that the mediator is close to on-
shell, thereby maximising the DM annihilation cross-section and hence 
minimising the value of $Y_\alpha$ required to obtain the correct relic 
density.~\footnote{For this choice of $m_\Phi/m_\rho$, 
the DM annihilation cross-section during freeze-out will feature a modest Breit-Wigner enhancement. This results in a freeze-out cross-section of $\mathcal{O}(10\%)$ larger than the $s$-wave contribution in the non-relativistic limit presented in eq.~\ref{eq:scalarDM_swaveA}.  This will make very little difference to the relic density line presented in fig.~\ref{fig:CS_s_chann_Ya}.}
If the mediator mass $m_\Phi$ is increased such that $m_\Phi > 3 m_\rho$, 
the $Y_\alpha$ lines (red lines) satisfying correct relic density move 
upwards and shrink the allowed DM parameter space. In particular, the DM 
parameter space is closed for $m_\Phi/m_\rho \simeq 15-18$ 
with the exact value dictated by the neutrino flavour to which the 
sterile state dominantly couples. The gray-shaded regions in both 
panels indicate the parameter space where the DM is overproduced.

\subsubsection{Existing constraints and future prospects}
\label{sssec:scalar_results}

In fig.~\ref{fig:CS_s_chann_Ya} we also present the existing constraints in the $Y_\alpha-m_\rho$ plane arising from indirect searches of $\rho\rho\rightarrow \nu\nu$ at various neutrino telescopes, from the non-unitarity of the light neutrino mixing matrix (denoted by EWPO hereafter), and from $N_{\rm{eff}}$ constraints on the dark matter mass.

The constraints from EWPO set upper limits on the 
active-sterile neutrino mixing $|U_{\alpha 4}|$. 
We represent these constraints in terms of the 
maximal values of $Y_\alpha$, denoted by the 
horizontal dotted blue lines (marked as $Y_\alpha^\textrm{MAX}$(EWPO) with $\alpha = e$ or $\mu$) in both panels of fig.~\ref{fig:CS_s_chann_Ya}. 
These lines correspond to the parameter choice 
$\mu_{\rho \phi}/m_\rho = 10$, $\lambda_{S \phi} = 1$, along with 
$|U_{e 4}|= 0.047$ and $|U_{\mu 4}|= 0.030$~\cite{Bolton:2022pyf}, respectively. The allowed $Y_\alpha^\textrm{MAX}$ would be increased by larger values of $\lambda_{S \phi}$, up to $\lambda_{S \phi} \sim 4\pi$, or larger values of the ratio $\mu_{\rho \phi}/m_\rho$. This would open up a larger region of viable parameter space, which extends to even higher values of the dark matter mass.
Note that the $Y_\alpha^\textrm{MAX}$ limits do not change  with the increase in the sterile neutrino  mass, since EWPO constraints are nearly constant for sterile neutrino masses $\gtrsim 100\,{\rm GeV}$. On the other hand, if the sterile neutrino mass is decreased, the constraints on $|U_{\alpha 4}|$ strengthen (see ~\cite{Bolton:2022pyf}) and hence $Y^{\rm MAX}_\alpha$ lines move downwards, thereby squeezing the allowed parameter space further.
 
Next we consider the constraints from indirect detection searches for $\rho\rho\rightarrow \nu\nu$ annihilation in the galaxy. In principle, these constraints depend on which neutrino flavour is produced in the final state, and therefore on which active flavour has the largest mixing with the sterile state. However, in practice, neutrino oscillations lead to appreciable fluxes of all three flavours, washing out the flavour dependence of the constraints.  The best limits then come from interactions involving electron flavour neutrinos. There are two existing constraints from the Super-Kamiokande (SK) experiment: the first, arising from a reinterpretation of the SK diffuse supernova neutrino background search (DSNB)~\cite{Olivares-DelCampo:2017feq}, is represented by the dashed green line labelled as SK~(DSNB) in the left panel of fig.~\ref{fig:CS_s_chann_Ya}. The second was obtained from a search for the relic supernovae electron antineutrino flux~\cite{Arguelles:2019ouk}, and is shown by the dashed purple line labelled as SK-$\overline{\nu_e}$ in the same panel. The latter constraint reaches the thermal relic cross section for a small mass range near 30 MeV. 

Big Bang Nucleosynthesis (BBN) constraints rule out complex scalar DM below $8.5\,$MeV~\cite{Sabti:2021reh}. This is represented by the light blue shaded vertical regions in fig.~\ref{fig:CS_s_chann_Ya}. 
Note that, the scalar mediator $\Phi$ does not mix with the SM Higgs due to the unbroken global $U(1)_L$, and hence, we only need to ensure that it decays before BBN. In fact, over the entire parameter space we have considered here, the total decay rate of $\Phi$: 
\begin{equation}
\Gamma_\Phi = \frac{\lambda^2_{S\phi}}{16\pi} \left(\sum_\alpha |U_{\alpha 4}|^2 \right)^2 m_\Phi + \frac{\mu^2_{\rho\phi}}{16\pi m_\Phi}\sqrt{1-\frac{4m^2_\rho}{m^2_\Phi}},  
\end{equation}
is sufficiently large to satisfy the BBN constraints. 

Finally, direct detection experiments do not impose any constraints on this model. This is because the $\rho\rho^*Z$ coupling vanishes in the limit of zero external momentum, and the tree-level $\rho\rho^*h$ coupling is tiny.

Future observational data will provide tighter constraints on this model as shown in the right panel of fig.~\ref{fig:CS_s_chann_Ya}. Indirect detection constraints from Hyper-Kamiokande~\cite{Bell:2020rkw} is projected to reach the thermal relic cross section for DM masses in the range $18 \MeV \lesssim m_\rho \lesssim 53\MeV$, shown by the dotdashed orange curve. 
We have also shown the projected sensitivity of JUNO~\cite{JUNO:2023vyz} by light green dotdashed line. The BBN constraints will also be tightened by the future CMB observations, with the projected constraints from CMB-S4 data indicated by the dotdashed brown line~\cite{Sabti:2019mhn}.

\subsection{Fermionic Dark Matter}
\label{ssec:fermion_DM}

Now we consider a model where the DM candidate is a SM singlet left-chiral Weyl fermion $\chi_L$. This model is a straightforward extension of the scalar DM model discussed in subsection~\ref{ssec:scalar_DM}. Keeping everything else intact, we now assume that the complex scalar $\rho$ serves as a mediator for DM annihilation to neutrinos, rather than being the DM itself. In order to achieve this, we assign a $U(1)_L$ charge of $+1/2$ to $\chi_L$; see Table~\ref{tab:ferm_qnums}. 

\begin{table}[htb!] 
    \centering
    \begin{tabular}{|c|c c c c c|}
    \hline
                 & $\chi_L$ & $\rho$ & $\Phi$ & $N_R$  & $S_R$\\\hline
         $U(1)_L$        & $+1/2$ & $+1$   & $+2$   & $+1$ & $-1$ \\
         \hline
    \end{tabular}
    \caption{Charges of the SM singlet scalars and fermionic fields under $U(1)_L$ in the fermionic DM model.} 
\label{tab:ferm_qnums}
\end{table}

\subsubsection{Overview of the model}
\label{sssec:ferm_overview}

The Lagrangian for this model is    
\begin{align}
    -\mathcal{L} \supset - \mathcal{L}_\nu + & \bigg(\frac{1}{2}m_\chi \overline{\chi_L}\chi^c_L + \frac{\lambda_{\chi \rho}}{2}\overline{\chi_L}\chi^c_L \rho +  \frac{\lambda_{S\phi}}{2}\overline{S^c_R}S_R\,\Phi \nonumber\\ 
    & + \frac{\lambda_{N\phi}}{2}\overline{N_R}N^c_R\Phi + h.c.\bigg)+ \tilde{V}(H,\Phi,\rho),
    \label{eq:lag_fermion}
\end{align}    
where the scalar potential, $\tilde{V}(H,\Phi,\rho)$, is now given by
\begin{equation}
\tilde{V}(H,\Phi,\rho) = V(H,\Phi,\rho) + \tilde{\mu}^2_{\rho\phi}\left(\rho\,\Phi^* + h.c. \right),
\label{eq:pot_fermion}
\end{equation}
and $V(H,\Phi,\rho)$ is same as in eq.~\ref{eq:pot_scalar}. 
As with the previous model, we again assume the quartic couplings $g_{\rho H}, g_{\phi H}$ and $g_{\rho \phi}$ to be small.

In this model, the global $U(1)_L$ is now {\it softly broken} by the $\Delta L = 1$ terms: $\frac{1}{2}m_\chi \left(\overline{\chi_L}\chi^c_L + h.c.\right)$ and $\tilde{\mu}^2_{\rho\phi}\left(\rho\,\Phi^* + h.c. \right)$. 
The purpose of adding these terms is two-fold: (i) to provide mass to the DM, and (ii) to allow mixing between $\rho$ and $\Phi$, such that the DM annihilation to neutrinos can take place via $s$-channel pseudoscalar mediated process. Note that, even after adding these $\Delta L = 1$ terms, there exists a remnant $Z_2$ symmetry which ensures the stability of the DM $\chi$.

\begin{figure}[htb!]
    \centering
    \includegraphics[width=0.38\textwidth]{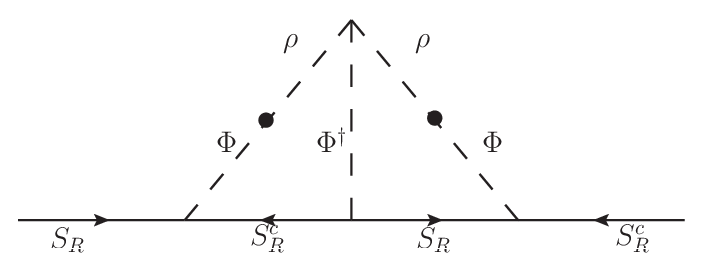}
    \caption{Feynman diagram generating a $\Delta L =2$ mass term for $S_R$ in the fermionic DM model. The circular blobs indicate the insertion of the $\Delta L=1$ terms present in the scalar potential $V(H,\Phi,\rho)$.} 
    \label{fig:majomass}
\end{figure}

Another important aspect of this model is that the $\Delta L = 1$ terms inevitably introduce the $\Delta L = 2$ mass term $\frac{1}{2}m_S \left(\overline{S_R^c}S_R + h.c.\right)$ for $S_R$, at the two-loop level; see fig.~\ref{fig:majomass}. The magnitude of this term is given by  
\begin{eqnarray}
m_S \propto \frac{\lambda^3_{S \phi} \sin^2\theta}{\left( 16\pi^2\right)^2}\,\mu_{\rho \phi} 
\end{eqnarray}
which is of $\mathcal{O}(1\,{\rm keV})$ for $\lambda_{S \phi} \sim 1, \sin\theta \sim 1$ and $\mu_{\rho\phi} \sim 20\,\MeV$, such that light neutrino masses of order $m_\nu \sim m_S\,\theta^2_\alpha \sim 0.1\,{\rm eV}$ would be obtained by setting $\theta_\alpha \sim 0.01$. However, we take $\mu_{\rho\phi} \ll 20\,\MeV$, so that our calculation in Appendix~\ref{app:Neutrinos} holds.
 
From the Lagrangian in eq.~\ref{eq:lag_fermion}, it is clear that our DM is a Majorana field, $\chi = \chi_L + \chi^c_L$. Let us express the two complex scalars in terms of real scalar and pseudoscalar fields as
\begin{align}
\Phi &= \phi+i \eta, \\
\rho &= \varrho + i a.
\end{align}
The presence of the $\tilde{\mu}^2_{\rho\phi}$ term in the scalar potential allows mixing between $\Phi$ and $\rho$, in both the scalar and the pseudoscalar sectors. For $s$-wave annihilation of the fermionic DM $\chi$, we are interested in the pseudoscalar mediators, for which the mass eigenstates are given by
\begin{align}
A_1 &= \cos\theta \eta + \sin\theta a ,\\
A_2 &= -\sin\theta \eta + \cos\theta a, 
\end{align}
with $m^2_{A_{1,2}} = (\mu^2_\phi+\mu^2_\rho) \pm (\mu^2_\phi-\mu^2_\rho)\sqrt{1+\tan^2 2\theta}$ 
and $\tan 2\theta = 2\tilde{\mu}^2_{\rho\phi}/\left(\mu^2_{\phi}-\mu^2_{\rho}\right)$.

\subsubsection{DM annihilation cross-sections and relic density}
\label{sssec:fermion_relic_abundance}

\begin{figure*}[thb!]
    \centering
    \includegraphics[width=8.9cm,height=6.0cm]{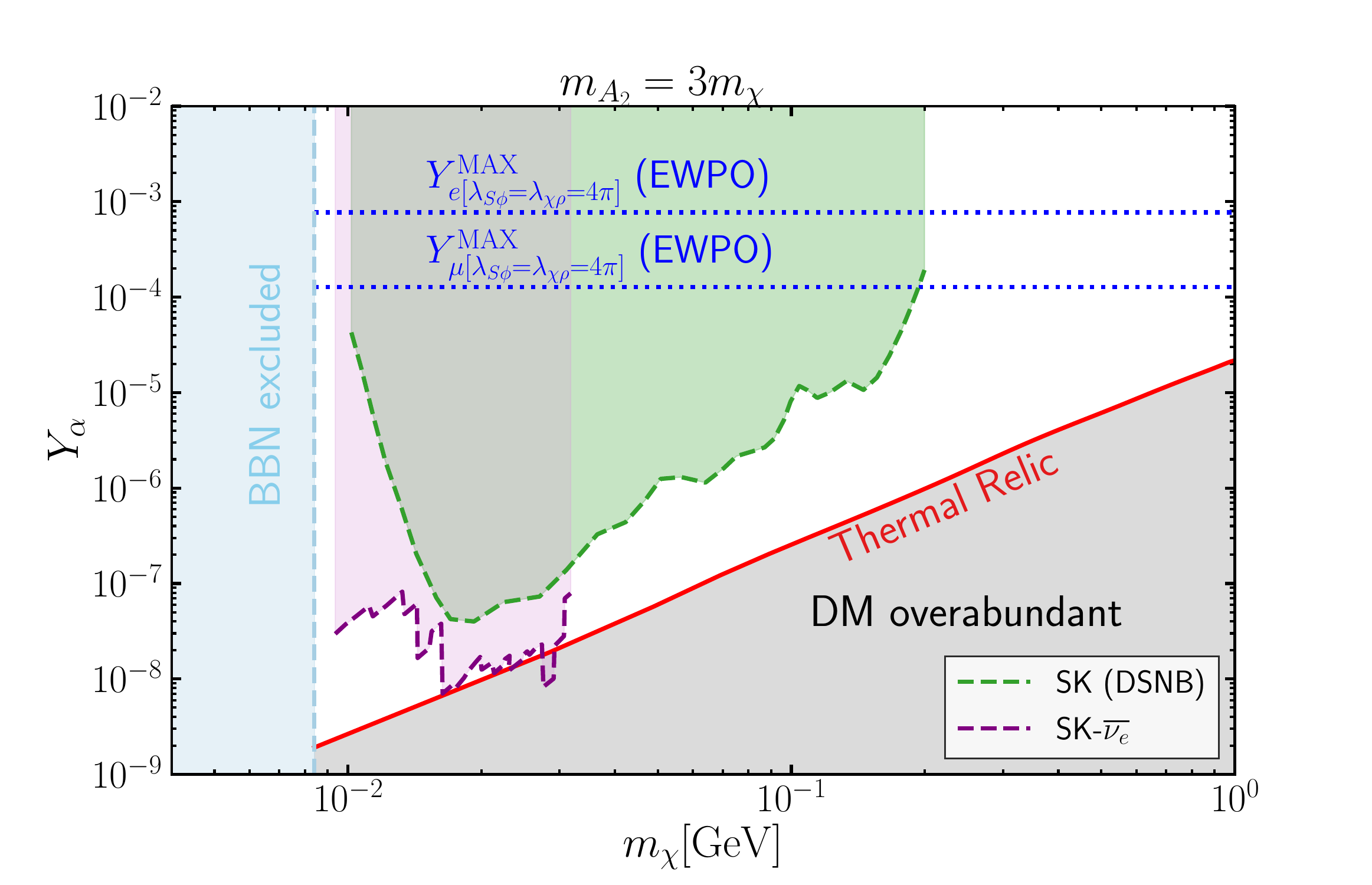}
    \includegraphics[width=8.9cm,height=6.0cm]{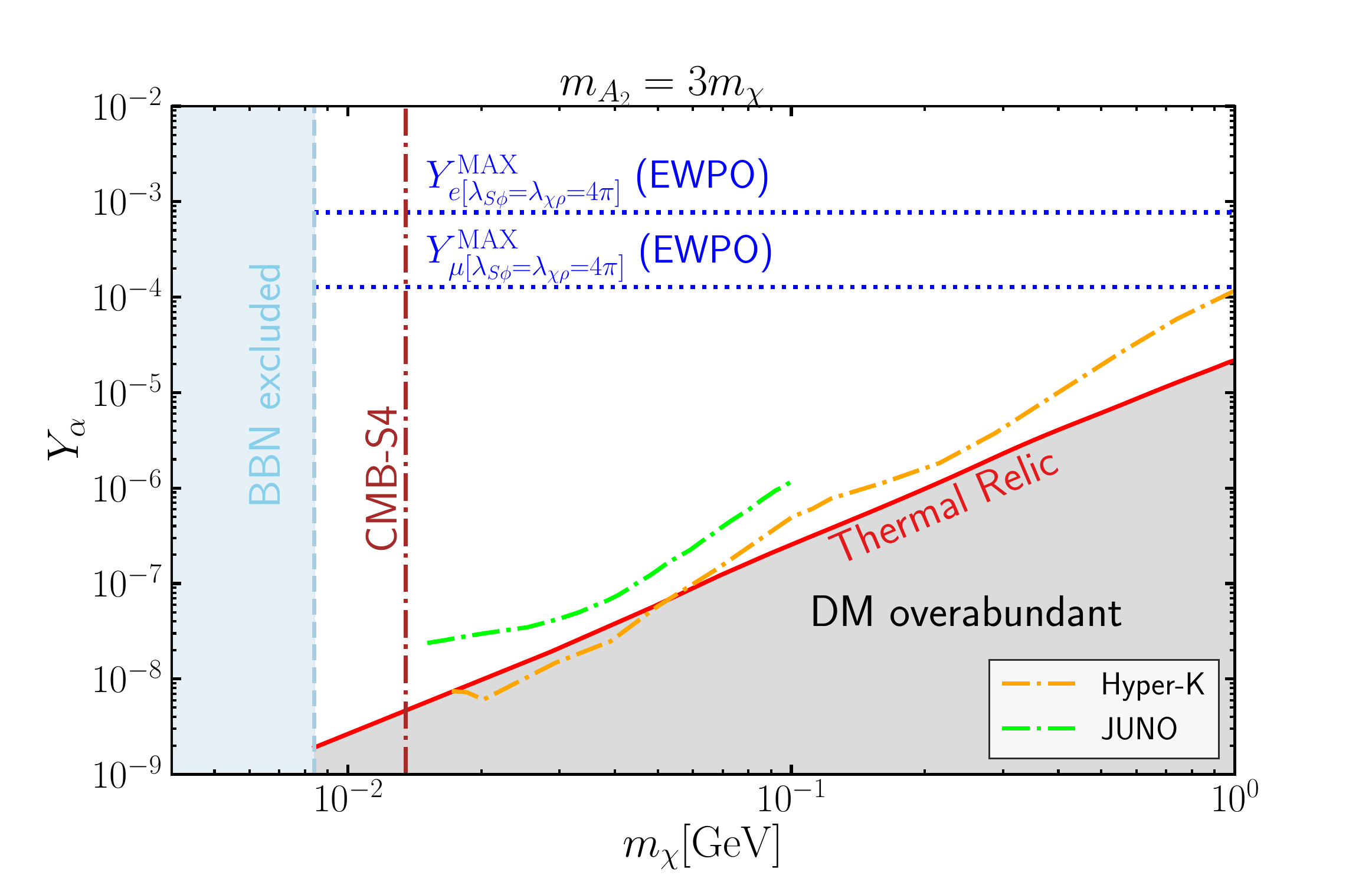}
    \caption{Parameter space for the Fermionic DM model with $m_{A_2}= 3 m_\chi$. The red lines show values of $Y_\alpha$ which lead to the correct DM relic abundance. The gray-shaded regions below these lines show where the DM is overproduced. The green and purple dashed lines in the left panels represent the existing constraints coming from Super-K, while the light blue shaded regions indicate BBN constraints. In the right panels, the projected sensitivities of the Hyper-K (orange) and JUNO (light green) are shown along with the CMB-S4 projected sensitivity (vertical brown dot-dashed line). The maximum allowed values of $Y_\alpha (\alpha=e,\mu)$ are shown by the horizontal blue dotted lines, assuming the parameter choice $\lambda_{S\phi}=\lambda_{\chi\rho}=4\pi$. See the text for details.
    }
    \label{fig:D_s_chann_Ya}
\end{figure*}

The $s$-wave annihilation of the DM $\chi$ to SM neutrinos is mediated by the pseudoscalars $A_1$ and $A_2$~\footnote{Note that there exist Feynman diagrams for DM annihilation via the real-scalar mediators composed of $\phi$ and $\varrho$.  However, those contributions are $p$-wave suppressed and can be neglected for the relic density calculation.}.  The relevant interaction terms can be read directly from eq.~\ref{eq:lag_fermion}. The DM-mediator couplings are given by
\begin{eqnarray}
\frac{\lambda_{\chi \rho}}{2}\overline{\chi_L}\chi^c_L \rho + h.c. \supset \frac{i \lambda_{\chi\rho}}{2} \overline{\chi}\gamma_5 \chi \left(\sin \theta A_1 + \cos \theta A_2 \right),\nonumber\\    
\label{eq:DMpseudoscalarint}
\end{eqnarray}
while neutrino-mediator couplings (see appendix~\ref{app:Neutrinos}) are
\begin{eqnarray}
&& \frac{\lambda_{S\phi}}{2}\overline{S^c_R}S_R\,\Phi + h.c.\nonumber\\ 
&& \supset i\lambda_{S\phi}\,\sum_\alpha |U_{\alpha 4}|^2 \overline{n_3}\gamma_5 n_3 \left(\cos \theta A_1 - \sin \theta A_2 \right). 
\label{eq:Nupseudoscalarint}
\end{eqnarray}
It is thus evident that there exists two pseudoscalar mediated Feynman diagrams for $\chi \chi \rightarrow \nu \nu$, involving $A_{1}$ and $A_{2}$.

For simplicity, we assume that $\tilde{\mu}^2_{\rho\phi} \ll \mu^2_{\rho} \ll \mu^2_{\phi}$, which implies that $m^2_{A_1} \gg m^2_{A_2}$.  In this limit, the $A_2$-mediated diagram makes the dominant contribution to the DM annihilation cross-section, with the other diagram being suppressed by the mass of heavier $A_1$. The relevant contribution to the annihilation cross-section is thus
\begin{align}
    \sigma_{ann} (\chi \chi \rightarrow \nu \nu) = & \frac{(\lambda_{\chi\rho}\lambda_{S\phi}\sin 2\theta)^2}{512\pi}\left( \sum_{\alpha} |U_{\alpha 4}|^2\right)^2 \nonumber\\
    & \times \frac{s}{(s - m^2_{A_2})^2}\sqrt{\frac{s}{s - 4\mchi^2}},
\end{align}
which has an $s$-wave piece given by
\begin{eqnarray}
\sigmav(\chi\chi \rightarrow \nu\nu) &=& \frac{(\lambda_{\chi\rho}\lambda_{S\phi}\sin 2\theta)^2}{64\pi}\left( \sum_{\alpha} |U_{\alpha 4}|^2\right)^2\nonumber\\
&& \times \frac{m^2_\chi}{\left(m^2_{A_2} - 4m^2_\chi\right)^2}.
\end{eqnarray}

Note that, unlike the scalar DM model, here the fermionic DM does not directly couple to the SM Higgs, and hence DM annihilation to light charged leptons is extremely suppressed.

For simplicity, we shall again assume that the sterile neutrinos mix with a single active flavour, such that the DM annihilation cross-section depends only on $m_\chi$, $m_{A_2}$ and the quantity 
\begin{equation}
Y_\alpha = (\lambda_{\chi\rho} \lambda_{S\phi} \sin 2\theta)^2 |\Uaf|^4. 
\end{equation}
We then have
\begin{eqnarray}
\sigmav(\chi\chi \rightarrow \nu_\alpha\nu_\alpha) &=& \frac{Y_\alpha}{64\pi} \frac{m^2_\chi}{\left(m^2_{A_2} - 4m^2_\chi\right)^2}.
\label{eq:fermionDM_swave}
\end{eqnarray}
Since the Majorana field $\chi$ is self-conjugate, this annihilation cross-section should be normalised to reproduce the entire relic density $\Omega_{\rm DM}$. 

Figure~\ref{fig:D_s_chann_Ya} shows the values of $Y_\alpha$ that give rise to the correct relic density (solid red lines), for $m_{A_2} = 3\,m_\chi$. As with the scalar model, if the mediator mass is increased, then larger values of $Y_\alpha$ are needed to achieve the thermal relic cross-section, which moves the thermal relic line upwards, reducing the available parameter space. 
Finally, the DM parameter space is completely closed when $m_{A_2}/m_\chi \simeq 20-30$, with the exact value depending on which neutrino flavour the sterile state dominantly couples to. The gray-shaded regions in both panels of fig.~\ref{fig:D_s_chann_Ya} indicate the parameter spaces where DM $\chi$ is overproduced. 


\subsubsection{Existing constraints and future prospects}
\label{sssec:fermion_results}

In fig.~\ref{fig:D_s_chann_Ya}, we compare the parameters required to achieve a thermal relic cross-section with the constraints arising from BBN, EWPO, and the results of indirect detection searches for DM annihilation to neutrinos.
The EWPO constraints, shown as the horizontal dotted blue lines, were obtained by using $\sin 2\theta = 1, \lambda_{\chi\rho} = \lambda_{S \phi} = 4\pi$ and setting $|U_{e 4}|= 0.047$ and $|U_{\mu 4}|= 0.030$~\cite{Bolton:2022pyf}, respectively. 
We also show the BBN constraint (vertical light blue shaded regions), the SK constraints arising from the reinterpretation of the DSNB search (green dashed line) and relic supernovae electron antineutrino search (purple dashed line) in the left panel of fig.~\ref{fig:D_s_chann_Ya}. 

In addition, in the right panel of 
fig.~\ref{fig:D_s_chann_Ya}, we have presented the 
projected sensitivities of some of the upcoming experiments 
by several dotdashed lines. The orange line represents 
the sensitivity of HyperK which will probe the DM 
relic line in the 18-53 MeV mass range. The 
light green dashed line represents the sensitivity 
of JUNO while the brown line corresponds to the 
projected sensitivity of CMB-S4.

\section{Summary and Conclusion}
\label{sec:conclusion}

Probes of the dark matter (DM) annihilation cross-section, via indirect detection, provide the most direct test of the thermal relic DM paradigm. With increasingly stringent indirect detection bounds on the annihilation of DM to visible final states, the annihilation to hard-to-detect final states, i.e., neutrinos, may be the last place for MeV-GeV scale thermal DM to hide (at least for $s$-wave annihilation). It is therefore imperative to think seriously about the neutrino-portal. In this work, we have expanded the range of viable models in which sub-GeV thermal DM annihilates dominantly into neutrinos.

We have proposed two neutrino-portal models, with either a scalar or fermionic DM candidate. In both cases, the DM is charged under an exact or softly broken global $U(1)_L$ symmetry which stabilizes the DM. In these models, a scalar or pseudoscalar field mediates interactions between the DM and a heavy SM singlet sterile neutrino state of mass $\sim 100\,{\rm GeV}$, which mixes with the light active neutrinos. In order to generate interestingly large active-sterile mixing angles of $\mathcal{O}(10^{-2})$, we assumed an inverse/linear see-saw like structure in the neutrino sector Lagrangian. Such large active-sterile mixing enables a substantial cross-section for DM annihilation to SM neutrinos, consistent with thermal relic DM. In both models, the annihilation of DM to charged leptons is negligible. 

Much of the allowed parameter space in these models will be explored by upcoming neutrino measurements at HyperKamiokande and CMB observations with CMB-S4. We have illustrated part of the viable parameter by fixing the relationship between DM and mediator mass, and also $\mu_{\rho\phi}$ and $m_{\rho}$ (in the first model) or $\lambda_{S\phi}$ and $\lambda_{\chi\rho}$ (in the second model). In future work, it would be interesting to explore the full parameter space more systematically via a global fit.

\section*{Acknowledgements}
This work was supported by the Australian Research Council through Discovery Project DP220101727. A.G. was supported by the ARC Centre of Excellence for Dark Matter Particle Physics, CE200100008.


\bigskip

\appendix
\section{Neutrino mixing details}
\label{app:Neutrinos}

After EWSB, the Lagrangian in eq.~\ref{eq:lagnu} gives rise to the 
following mass terms in the neutrino sector:  
\begin{equation}
\footnotesize
    \frac{1}{2}\begin{pmatrix} \overline{\nu_{e L}} \\ \overline{\nu_{\mu L}} \\\overline{\nu_{\tau L}} \\ \overline{S^c_R} \\ \overline{N^c_R} \end{pmatrix}^T \underset{M}{\underbrace{\begin{pmatrix}
    0 & 0 & 0 & 0 & \theta_{e} m_N \\
    0 & 0 & 0 & 0 & \theta_{\mu} m_N \\
    0 & 0 & 0 & 0 & \theta_{\tau} m_N \\
    0 & 0 & 0 & 0 & m_N \\
    \theta_{e} m_N & \theta_{\mu} m_N & \theta_{\tau} m_N & m_N & 0
    \end{pmatrix}}}
\begin{pmatrix}
\nu_{eL}^c \\ \nu_{\mu L}^c \\ \nu_{\tau L}^c\\ S_R \\ N_R
\end{pmatrix}
 + h.c.
\end{equation}
where $\nu_{\alpha L}^c\equiv (\nu_{\alpha L})^c$ and 
$\theta_\alpha = y_{\nu\alpha} v/\sqrt{2} m_N$. 

The physical mass eigenstates in the neutrino sector 
are obtained when the mass matrix $M$ is diagonalized 
as follows:  
\begin{equation}
    U^T M U = \text{diag}(m_1, m_2, m_3, m_4, m_5) = M^d
\end{equation}
with all $m_i$ real and positive. As we are only 
considering real values of $y_{\nu \alpha}$ and $m_N$, 
$M$ is real and symmetric, and so it can be diagonalised 
by its eigenvalue decomposition along with the 
inclusion of a phase matrix 
$P = \text{diag}(1, 1, 1, 1, i)$, so 
that the eigenvalues $\{m_i\}$ are positive. A Gram-Schmidt orthogonalization leads to the following orthogonal mixing matrix:
\onecolumngrid
\begin{align}
    U & = \begin{pmatrix} 
    -\frac{\theta_\mu}{\sqrt{\theta_e^2 + \theta_\mu^2}} & -\frac{\theta_e \theta_\tau}{\sqrt{(\theta_e^2 + \theta_\mu^2)\sum\nolimits_{\alpha} \theta_\alpha^2}} & -\frac{\theta_e}{\sqrt{(1+ \sum\nolimits_{\alpha} \theta_\alpha^2) \sum\nolimits_{\alpha} \theta_\alpha^2}} & \frac{\theta_e}{\sqrt{2}\sqrt{1 + \sum\nolimits_{\alpha} \theta_\alpha^2}} & -\frac{i \theta_e}{\sqrt{2}\sqrt{1 + \sum\nolimits_{\alpha} \theta_\alpha^2}} \\
    \frac{\theta_e}{\sqrt{\theta_e^2 + \theta_\mu^2}} & -\frac{\theta_\mu\theta_\tau}{\sqrt{(\theta_e^2 + \theta_\mu^2) \sum\nolimits_{\alpha} \theta_\alpha^2}} & -\frac{\theta_\mu}{\sqrt{(1+\sum\nolimits_{\alpha} \theta_\alpha^2) \sum\nolimits_{\alpha} \theta_\alpha^2}} & \frac{\theta_\mu}{\sqrt{2}\sqrt{1 + \sum\nolimits_{\alpha} \theta_\alpha^2}}& -\frac{i\theta_\mu}{\sqrt{2}\sqrt{1 + \sum\nolimits_{\alpha} \theta_\alpha^2}}\\
    0 & \sqrt{\frac{\theta_e^2 + \theta_\mu^2}{\sum\nolimits_{\alpha} \theta_\alpha^2}} & -\frac{\theta_\tau}{\sqrt{(1+ \sum\nolimits_{\alpha} \theta_\alpha^2) \sum\nolimits_{\alpha} \theta_\alpha^2}} & \frac{\theta_\tau}{\sqrt{2}\sqrt{1 + \sum\nolimits_{\alpha} \theta_\alpha^2}} & -\frac{i\theta_\tau}{\sqrt{2}\sqrt{1 + \sum\nolimits_{\alpha} \theta_\alpha^2}} \\ 
    0 & 0 & \sqrt{\frac{\sum\nolimits_{\alpha} \theta_\alpha^2}{1+\sum\nolimits_{\alpha} \theta_\alpha^2}} & \frac{1}{\sqrt{2}\sqrt{1 + \sum\nolimits_{\alpha} \theta_\alpha^2}}  &  -\frac{i}{\sqrt{2}\sqrt{1 + \sum\nolimits_{\alpha} \theta_\alpha^2}}  \\
    0 & 0 & 0 & \frac{1}{\sqrt{2}} & \frac{i}{\sqrt{2}}
    \end{pmatrix}.
\label{eq:Umat}
\end{align}
\twocolumngrid

The flavour eigenstates are now related to the mass eigenstates as
\begin{align}
    \tilde{\nu}_{L,R} & = U\tilde{n}_{L,R}
    \label{eq:masstransform}
\end{align}
where 
$\tilde{\nu}_{L} = (\nu_{e L}, \nu_{\mu L}, \nu_{\tau L}, S^c_R, N^c_R)$
and $\tilde{n}_{L} = (n_{L 1}, n_{L 2}, n_{L 3}, n_{L 4}, n_{L 5})$. 
In the diagonal basis, the mass terms in the Lagrangian become 
\begin{equation}
    \frac{1}{2}\overline{(n_{L}^c)_i} M^d_{ij}(n_{L})_j +h.c. = \frac{1}{2}\sum_{i=1}^5 m_i \overline{(n_{L}^c)_i}(n_{L})_j + h.c.
\end{equation}

Let us now define a left-handed field 
$\mathcal{N}_L = U_{\alpha 4}\nu_{\alpha L} + U_{S 4} S_R^c$. 
Using the fact that $U_{\alpha 5} = -i U_{\alpha 4}$ for 
$\alpha \neq N$ and $U_{N5} = i U_{N4}$ (see~\ref{eq:Umat}), we can write
\begin{align}
    (n_{L})_4 &= \mathcal{N}_L + U_{N4} N_R^c, \\
    (n_{L})_5 & = i\mathcal{N}_L  - i U_{N4} N_R^c,
\end{align}
such that the corresponding mass terms can be written as, 
\begin{eqnarray}
&& \frac{1}{2} m_4 \left((\overline{n^c_{L}})_4 (n_{L})_4 + (\overline{n^c_{L}})_5 (n_{L})_5 \right) + h.c. \nonumber\\
& =& \sqrt{2} m_4 \left( \overline{N_R} \mathcal{N}_L +\overline{\mathcal{N}_L} N_R\right) 
\label{eq:masstermsfinal}
\end{eqnarray}
where $m_1 = m_2 = m_3 = 0$ and 
$m_4 = m_5 = m_N \sqrt{1+ \sum\nolimits_{\alpha} \theta_\alpha^2}$.  We thus have 3 massless `light' neutrinos 
and a pair of degenerate Majorana neutrinos that form a Dirac neutrino. Defining
\begin{equation}
    n_D = \begin{pmatrix}
    U_{\alpha 4} \nu_{\alpha L} + U_{S 4} S^c_R\\ N_R
    \end{pmatrix}
\end{equation}
we see that (from eq.~\ref{eq:masstermsfinal}) 
\begin{equation}
\sqrt{2} m_4 \left( \overline{N_R} \mathcal{N}_L +\overline{\mathcal{N}_L} N_R\right) \rightarrow \sqrt{2} m_4 \overline{n_D} n_D.
\end{equation}
Now, from eq.~\ref{eq:masstransform} one can write, 
\begin{align}
    S^c_R &= \sum_i U_{Si} (n_{L})_i + 2 U_{S4} (n_{D})_L,\\
    S_R &= \sum_i U_{Si}^* (n_{R})_i + 2 U_{S4}^* (n_{D})_R,\\
    N_R &= (n_{D})_R.
\label{eq:sterile-active_mix}    
\end{align}
where $(n_{R})_i = (n_{L}^c)_i$ and $i$ runs from 1 to 3.

In  both the models we have considered, the 
mediator-neutrino coupling is given by,
\begin{eqnarray}
\frac{\lambda_{S\phi}}{2}\overline{S^c_R}S_R\,\Phi + h.c. &\rightarrow & \frac{\lambda_{S\phi}}{2}\,|U_{S 3}|^2\left(\overline{n_3}n_3\phi + i\,\overline{n_3}\gamma_5 n_3 \eta \right)
\nonumber\\
& \rightarrow & \lambda_{S\phi}\,\sum_\alpha |U_{\alpha 4}|^2\left(\overline{n_3}n_3\phi + i\,\overline{n_3}\gamma_5 n_3 \eta \right),\nonumber\\
\label{eq:numediator_coupling}
\end{eqnarray}
where $\alpha = e,\mu,\tau$ and in the 
last line of eq.~\ref{eq:numediator_coupling} 
we have used 
$|U_{S 3}|^2 = 2\sum_\alpha |U_{\alpha 4}|^2$, 
as obtained from \ref{eq:Umat}. 

\section{Thermally averaged annihilation cross-sections}
\label{app:Annxsec_Thermav}

Given the matrix element $\mathcal{M}$ for the 
${\rm DM}\,{\rm DM} \rightarrow {\rm SM}\,{\rm SM}$ 
process, we calculate the annihilation cross-section 
as follows~\cite{1991NuPhB.360..145G}:
\begin{equation}
\sigma_{ann} = \frac{1}{2}\frac{\beta_f}{16\pi\,g^2_{\rm DM}\,\sqrt{s(s-4m^2_{\rm DM})}} \int \frac{d\cos\theta}{2}\frac{d\phi}{2\pi}|\mathcal{M}|^2,    
\label{eq:xsec}
\end{equation}
where, $\beta_f = \sqrt{1-\dfrac{4m^2_{\rm SM}}{s}}$, with 
$s$ being the invariant mass squared of the incoming 
two-particle DM system. In eq.~\ref{eq:xsec}, $g_{\rm DM}$ 
represents the degrees of freedom of the DM candidate, 
and the factor of $\frac{1}{2}$ is due to the production 
of identical final state particles. For obtaining 
the $s$-wave piece of the annihilation 
cross-section we have used~\cite{1991NuPhB.360..145G}:
\begin{equation}
\sigma_{ann}v = \frac{1}{2}\frac{\beta_f}{16\pi\,g^2_{\rm DM}\,(s-2\,m^2_{\rm DM})} \int \frac{d\cos\theta}{2}\frac{d\phi}{2\pi}|\mathcal{M}|^2,    
\label{eq:xsecv}
\end{equation}
and then expanded $(\sigma_{ann}v)$ in series of 
$\epsilon = \frac{s-4m^2_{\rm DM}}{4m^2_{\rm DM}}$, 
with the zeroth order term giving the required 
$s$-wave piece of the DM annihilation 
cross-section. 

\begin{enumerate}
\item \textbf{Scalar dark matter:} For the scalar dark 
matter model considered in Sec.~\ref{ssec:scalar_DM}, 
the matrix-element squared for DM annihilation 
process ($\rho \rho \rightarrow \nu \nu$) is:
\begin{equation}
|\mathcal{M}|^2 = \frac{\mu^2_{\rho \phi}}{4}\lambda^2_{S \phi}\left( \sum_\alpha |U_{\alpha 4}|^2\right)^2 \frac{4\,s}{(s-m^2_\Phi)^2},    
\end{equation}
Combining this with eq.~\ref{eq:xsec} 
and using $g_{\rm DM} =2, \beta_f = 1$ we 
obtain:
\begin{align}
\sigma_{ann} = \frac{(\mu_{\rho \phi } \lambda_{S\phi})^2 }{128\pi} & \left(\sum_\alpha |U_{\alpha 4}|^2\right)^2 \nonumber\\ 
     & \times \frac{1}{(s - m_\Phi^2)^2}\sqrt{\frac{s}{s - 4m_\rho^2}}.    
\end{align}

\item \textbf{Fermionic dark matter:} For the fermionic 
dark matter model considered in Sec.~\ref{ssec:fermion_DM}, 
the matrix-element squared for the DM annihilation process 
($\chi\chi \rightarrow \nu\nu$) 
is given by:
\begin{equation}
|\mathcal{M}|^2 = \frac{\lambda^2_{\chi\rho}}{4}\lambda^2_{S \phi}\sin^2 2\theta\left( \sum_\alpha |U_{\alpha 4}|^2\right)^2 \frac{s^2}{(s-m^2_{A_2})^2},    
\end{equation}
Combining this with eq.~\ref{eq:xsec} 
and using $g_{\rm DM} =2, \beta_f = 1$ we 
obtain:
\begin{align}
    \sigma_{ann}= \frac{(\lambda_{\chi\rho}\lambda_{S\phi}\sin 2\theta)^2}{512\pi} & \left( \sum_{\alpha} |U_{\alpha 4}|^2\right)^2 \nonumber\\
    & \times \sqrt{\frac{s}{s - 4\mchi^2}}\frac{s}{(s - m^2_{A_2})^2 }.
\end{align}

\end{enumerate}

\label{Bibliography}
\bibliography{refs} 
\end{document}